\documentclass[lettersize,journal]{IEEEtran}
\usepackage{graphicx}
\usepackage{setspace}
\usepackage{adjustbox} 
\DeclareGraphicsExtensions{.pdf,.jpeg,.png}
\usepackage{enumitem}
\usepackage[noadjust]{cite}
\usepackage[linesnumbered,ruled]{algorithm2e}
\usepackage{amsmath}
\usepackage{lipsum}
\usepackage{mathtools}
\usepackage{cuted}
\usepackage{array}
\usepackage{subfigure}
\usepackage{amssymb}
\usepackage{caption}
\usepackage{multirow}
\usepackage{makecell}
\usepackage{algorithmic}
\usepackage{xcolor}

\SetCommentSty{mycommfont}
\usepackage{array,multirow,makecell}
\SetKwInput{KwInput}{Input}    
\SetKwInput{KwRepeat}{Repeat} 
\SetKwInput{KwOutput}{Output}    

\begin{document}


\title{Gradient-Based Meta Learning for Uplink RSMA with Beyond Diagonal RIS}

\author{Shreya Khisa, Ali Amhaz, Mohamed Elhattab, Chadi Assi, and Sanaa Sharafeddine}

\maketitle

\begin{abstract} 
Beyond diagonal reconfigurable intelligent surface (BD-RIS) has emerged as an innovative and generalized RIS framework that provides greater flexibility in wave manipulation and enhanced coverage. In comparison to conventional RIS, optimization of BD-RIS is more challenging due to the large number of optimization variables associated with it. Typically, optimization of large-scale optimization problems utilizing traditional optimization methods results in high complexity. To tackle this issue, we propose a gradient-based meta learning algorithm which works without pre-training and is able to solve large-scale optimization problems. With the objective to maximize the sum rate of the system, to the best of our knowledge, this is the first work considering joint optimization of receiving beamforming vectors at the base station (BS), scattering matrix of BD-RIS and transmission power of users equipment (UEs) in uplink rate-splitting multiple access (RSMA) communication. Numerical results demonstrate that our proposed scheme can outperform the conventional RIS RSMA framework by 22.5$\%$.

\end{abstract}


\section{Introduction}
The sixth generation (6G) of wireless communication is expected to provide 10 times more spectral efficiency and 100 times more energy efficiency \cite{10054381}. To fulfill such goals,  reconfigurable intelligent surface (RIS) has been envisioned as a promising enabling technology for next-generation wireless communications. An RIS is a two-dimensional planar surface consisting of numerous passive/active elements. Each element/meta-atom can control wireless propagation channels, and thus, can enhance the spectral efficiency of the wireless network. The benefits of utilizing RIS in a wireless environment have already been demonstrated in many wireless systems \cite{9424177}.

\par Recently, beyond diagonal RIS (BD-RIS), which is referred to as RIS 2.0, has emerged as a promising technology for next-generation wireless networks \cite{9913356, 10316535}. Depending on its structure, BD-RIS can be classified into single, group, and fully connected architectures. The conventional RIS, also known as RIS 1.0, essentially has the single-connected architecture and works under the reflective mode, which mathematically yields a diagonal phase shift matrix \cite{9913356}. It has a simpler structure but with limited performance. This type of RIS has been widely investigated by the research community \cite{10109654}. On the other hand, a fully connected architecture offers the highest flexibility and optimization potential. However, it suffers from high computational complexity and control requirements \cite{10316535}. Meanwhile, the group-connected architecture divides the RIS elements into groups and can also provide high performance while keeping the complexity at a moderate level \cite{10316535}. 
\par Meanwhile, rate-splitting multiple access (RSMA) has recently been widely accepted as one of the potential candidates for next-generation multiple access (NGMA) techniques for 6G wireless networks \cite{10109654}. RSMA offers a more versatile and resilient transmission framework compared to space division multiple access (SDMA) and non-orthogonal multiple access (NOMA) \cite{10109654}. The main working principle behind RSMA is that it partially treats user interference as noise and partially decodes the interference. This is achieved by splitting the messages of users equipment (UEs) into common and private parts and transmitting them using superposition coding (SC). In this transmission scheme, the common message can be decoded by multiple UEs and subsequently removed from the received signal using successive interference cancellation (SIC). Meanwhile, each private message is then decoded solely by its intended UE.
\par Research efforts on BD-RIS are still in their early stages. The authors in \cite{10155675} proposed a passive beamforming optimization problem for BD-RIS to maximize the BD-RIS-assisted channel gains. However, this study only investigated a single-antenna base station (BS), and therefore, it will not apply to multi-antenna-based systems. The authors in \cite{10288244} proposed a BD-RIS-assisted RSMA framework for ultra-reliable low-latency communications (URLLC) systems, considering downlink transmission. In \cite{10364738}, the authors proposed a unified approach to optimize power consumption, energy efficiency, and sum rate for BD-RIS. The work in \cite{10411856} investigated the synergy between RSMA and BD-RIS to enhance network coverage, improve performance, and reduce the need for antenna elements. The above-mentioned studies have addressed these issues using various optimization methods, such as alternating optimization (AO), successive convex approximation (SCA), and semi-definite relaxation (SDR). However, these methods suffer from high complexity in large-scale scenarios, and their performance is not guaranteed in dynamic environments.
\par Recently, the use of deep learning (DL) in wireless communications has attracted significant attention due to its ability to address complex and challenging problems \cite{8847377,9834153}. However, these data-driven approaches fall behind in dynamic scenarios, as most of them require comprehensive training datasets covering all scenarios, which is quite rare to find. Very recently, meta-learning-based approaches have been getting enormous attention from the research community to solve large-scale optimization problems due to their capability to adapt to dynamic environments and provide low-complexity solutions. \cite{10623434,zou2021meta,wang2024energy,eghbali2024providing,loli2024meta,amiri2024multi}. The authors in \cite{10623434} studied a gradient and manifold-based meta-learning solution to optimize the phase shift matrix of RIS and beamforming vectors at BS. Meanwhile, the authors in \cite{zou2021meta} proposed a meta-learning-aided solution for RIS-aided NOMA networks. The authors in \cite{eghbali2024providing} investigated a system combining multiple simultaneously transmitting and reflecting RIS (STAR-RIS) with full-duplex (FD) communication, aimed at providing URLLC services utilizing meta-learning. The authors in \cite{loli2024meta} have shown the effectiveness of gradient-based meta-learning in solving a large-scale optimization problem.
\par Although several studies in the literature have addressed BD-RIS and BD-RIS with RSMA in downlink communication \cite{10155675}-\cite{10411856}, no work has investigated the performance of BD-RIS or BD-RIS with RSMA in uplink. Motivated by this gap, and to the best of our knowledge, this is the first paper to study the performance of BD-RIS in uplink, along with the integration of RSMA. Additionally, inspired by the superior performance of gradient-based meta-learning in solving large-scale optimization problems, we employ this method to solve our formulated optimization problem.
\section{System Model}
\subsection{BD-RIS architecture}
An $M$-element RIS can be imagined as $M$ antennas connected to an $M$ port scattering network \cite{10316535}. Particularly, let us define the incident signal as $\textbf{x} \in \mathbb{C}^{M \times 1}$ and reflected signal as  $\textbf{y} \in \mathbb{C}^{M \times 1}$ of the $M$ ports respectively. Then, following the scattering network theory, the input-output relation of this $M$-port network can be represented as 
\begin{equation}
    \textbf{y}=\boldsymbol{\Phi}\textbf{x},
\end{equation}
where the reflection coefficient matrix is represented by $\boldsymbol{\Phi} \in \mathbb{C}^{M \times M}$.
There are three existing types of BD-RIS architecture:  single-connected, fully-connected, and group-connected BD-RIS. 

\textbf{Single-connected BD-RIS}: In a single-connected BD-RIS architecture, there is no connection between the ports. RIS 1.0 falls under this category. The single-connected RIS should satisfy the following constraint,
\begin{equation}
    \mathcal{B}_1=\{\boldsymbol{\Phi}|\boldsymbol{\Phi}=\textrm {diag}(e^{j\theta_1},e^{j\theta_2},....e^{j\theta_M})\},
\end{equation}
where $\theta_M \in [0,2\pi)$ denotes the phase shift angle. 

\textbf{Fully-connected BD-RIS}: In a fully-connected BD-RIS architecture, each port is connected with all other ports through reconfigurable reactance. This can be seen as a generalized framework for all types of RIS. Consequently, the scattering matrix of fully-connected BD-RIS should satisfy the following constraint,
\begin{equation}
    \mathcal{B}_2=\{\boldsymbol{\Phi}| \boldsymbol{\Phi}=\boldsymbol{\Phi}^T,\boldsymbol{\Phi}\boldsymbol{\Phi}^H=\boldsymbol{I}\}.
\end{equation}

\textbf{Group-connected BD-RIS}: For the group connected BD-RIS, $M$ number of BD-RIS elements is divided into $G$ number of groups, where each group contains $M_g$ number of elements. Elements in the same group are connected to all other elements. Meanwhile, there is no connection between elements of different groups. Consequently, the scattering matrix must satisfy the following constraint,
\begin{equation}
    \mathcal{B}_3=\{\boldsymbol{\Phi}| \boldsymbol{\Phi}=\textrm {blkdiag}(\boldsymbol{\Phi}_1,....,\Phi_G), \boldsymbol{\Phi}_g^H\boldsymbol{\Phi}_g=\boldsymbol{I},\boldsymbol{\Phi}_g=\boldsymbol{\Phi}_g^T\},
\end{equation}
where $\boldsymbol{\Phi}_g, g \in {1,....,G}$ are complex symmetric unitary matrices and blkdiag(.) means the block diagonal matrix. It has been shown in many literature that group-connected BD-RIS can achieve a higher performance with moderate complexity \cite{9913356, 10316535}. Hence, in our proposed work we utilize group-connected BD-RIS.
\subsection{Transmission Model} 
\begin{figure}[!t]
    \centering
\includegraphics[width=0.9\columnwidth]{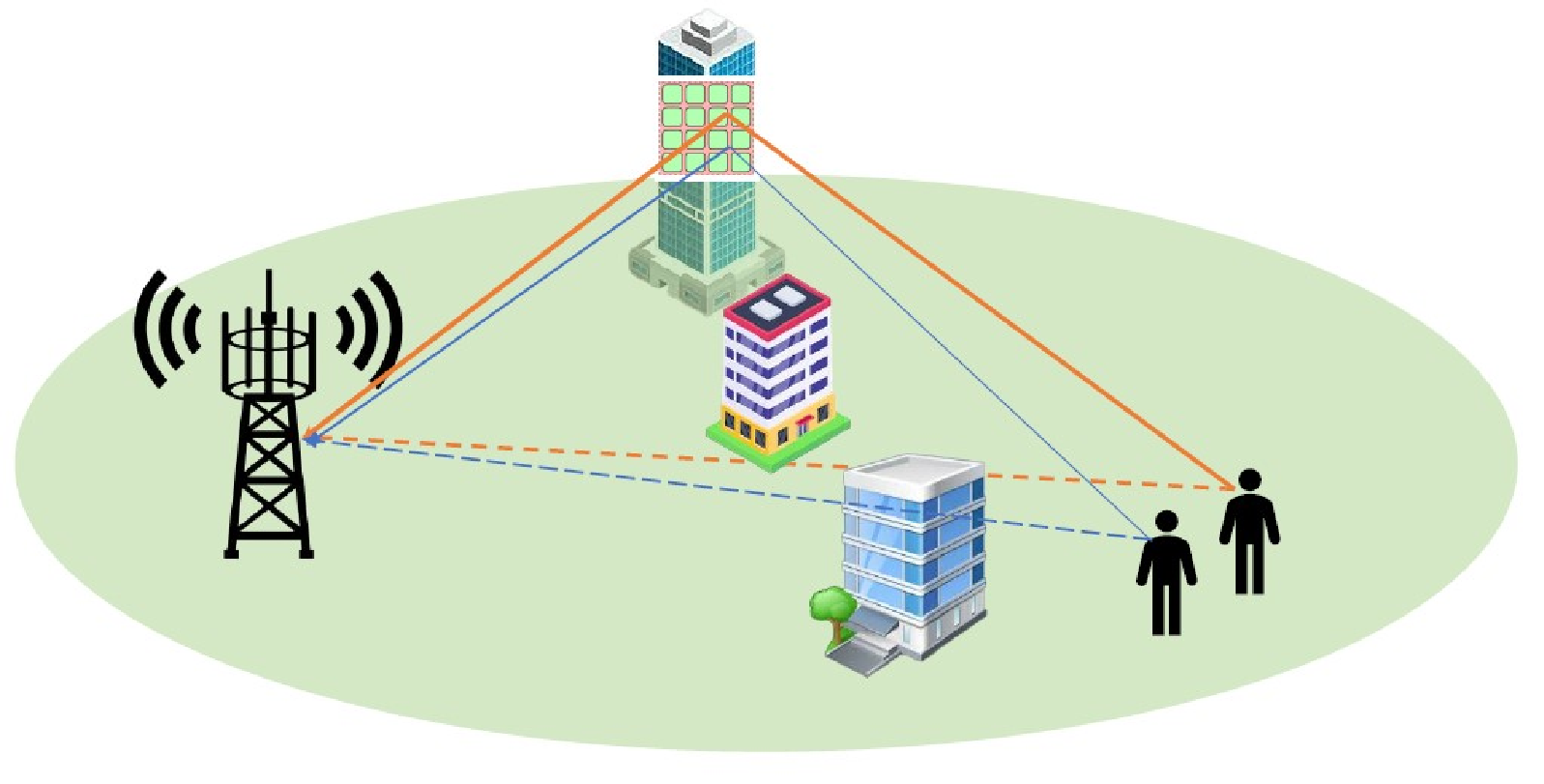}
\caption{BD-RIS-assisted RSMA in uplink communication}
\label{system_model}
\end{figure}

As shown in Fig. \ref{system_model}, we consider an uplink scenario where a BS is equipped with $N$ antennas, a group-connected BD-RIS consists of $M$ elements and two UEs equipped with single antennas. We assume that RSMA principle is applied as a MA technique for UEs to communicate with the BS. According to uplink RSMA principle, the message of UE-1, $W_1$, is split into two sub-messages, $W_{11}$ and $W_{12}$. Afterward, these two sub-messages are independently encoded into two streams $s_{11}$ and $s_{12}$. Meanwhile, the message of UE-2, $W_2$ is kept without splitting and this message goes through the encoding process independently which generates the stream $s_2$. It is proven in uplink RSMA that it is enough to split $K-1$ messages when there are $K$ UEs to achieve the capacity region \cite{9064705}. 
Furthermore, we assume an urban environment where obstacles and blockages weaken the direct communication path between the UEs and the BS. Hence, we utilize a BD-RIS, which can establish a line-of-sight (LoS) communication with the BS and assist the UEs in transmitting their signals with enhanced signal quality. The signals are transmitted through both the direct path from the UEs to the BS and the cascaded path via the UE-BD-RIS-BS link. We assume perfect knowledge of the channel state information (CSI). Thus, the transmitted signal from UE-1 to BS can be denoted as,
\begin{equation}
    x_1 = \sqrt{p_{11}}s_{11}+\sqrt{p_{12}}s_{12},
\end{equation}
where $p_{11}$ and ${p_{12}}$ represent the power allocation for stream $s_{11}$ and $s_{12}$. Meanwhile, the 
received signal at BS from UE-1 can be denoted by,
\begin{equation}
{\boldsymbol{y}_{b,1}=\left(\textbf{h}_{1,b}+\textbf{h}_{r,b}\boldsymbol{\Phi}\textbf{h}_{r,1}\right)\left(\sqrt{p_{11}}s_{11}+\sqrt{p_{12}}s_{12}\right)+ 
    \textbf{n}_{1},}\label{Sig1}
\end{equation}
where $\textbf{h}_{1,b} \in \mathbb{C}^{N \times 1}$ represents the channel coefficient between UE-1 and BS, $\textbf{h}_{r,b} \in \mathbb{C}^{N \times M}$ represents the channel coefficient between the BD-RIS and BS, $\boldsymbol{\Phi}$ represents the scattering matrix of BD-RIS and $\textbf{h}_{r,1} \in \mathbb{C}^{M \times 1}$ represents the channel coefficient between BD-RIS and UE-1 and $\textbf{n}_1 \in \mathbb{C}^{N \times 1}$ represents the additive white Gaussian noise with zero mean and variance of $\sigma^2$. On the other hand, the transmitted signal from UE-2 is denoted as,
\begin{equation}
    x_2 = \sqrt{p_{2}}s_{2},
\end{equation}
where $p_2$ represents the transmission power of UE-2 to transmit $s_2$. Thus, the received signal from UE-2 at BS can be denoted by, 
\begin{equation}
{\boldsymbol{y}_{b,2}=\left(\textbf{h}_{2,b}+\textbf{h}_{r,b}\boldsymbol{\Phi}\textbf{h}_{r,2}\right)\left(\sqrt{p_{2}}s_{2}\right)+ 
    \textbf{n}_{2},}\label{Sig1}
\end{equation}
where $\textbf{h}_{2,b} \in \mathbb{C}^{N \times 1}$ represents the channel coefficient between UE-2 and BS, $\textbf{h}_{r,b} \in \mathbb{C}^{N \times M}$ represents the channel coefficient between the BD-RIS and BS,  $\textbf{h}_{r,2} \in \mathbb{C}^{M \times 1}$ represents the channel coefficient between BD-RIS and UE-2, and $\textbf{n}_2 \in \mathbb{C}^{N \times 1}$ represents the additive white Gaussian noise. We assume that at BS, the streams are decoded in the following decoding order: $s_{11}, s_2, s_{12}$ \cite{10423227}.
Hence, the achievable rate to decode the stream $s_{11}$ of UE-1 at BS can be given by,
\begin{align}\footnotesize
&R_{11}(\textbf{P},\textbf{W},\boldsymbol{\Phi})=\log_2\left(1+ \right. \\ 
& \footnotesize\left.\frac{p_{11}|\textbf{w}_{11}^H\left(\textbf{h}_{1,b}+\textbf{h}_{r,b}\boldsymbol{\Phi}\textbf{h}_{r,1}\right)|^2}{p_{12}|\textbf{w}_{11}^H(\textbf{h}_{1,b}+\textbf{h}_{r,b}\boldsymbol{\Phi}\textbf{h}_{r,1})|^2+
p_{2}|\textbf{w}_{11}^H(\textbf{h}_{2,b}+\textbf{h}_{r,b}\boldsymbol{\Phi}\textbf{h}_{r,2})|^2+\sigma^2}\right),
\end{align}
where $\textbf{w}_{11} \in \mathbb{C}^{N \times 1}$ represents the receiving beamforming vector at the BS to decode stream $s_{11}$. Meanwhile, the achievable rate to decode the stream $s_{2}$ of UE-2 at BS can be given by,
\begin{equation}
R_{2}(\textbf{P},\textbf{W},\boldsymbol{\Phi})=\log_2\left(1+\frac{p_{2}|\textbf{w}_{2}^H\left(\textbf{h}_{2,b}+\textbf{h}_{r,b}\boldsymbol{\Phi}\textbf{h}_{r,2}\right)|^2}{p_{12}|\textbf{w}_{2}^H(\textbf{h}_{1,b}+\textbf{h}_{r,b}\boldsymbol{\Phi}\textbf{h}_{r,1}|^2)+\sigma^2}\right),
\end{equation}
where $\textbf{w}_{2} \in \mathbb{C}^{N \times 1}$ represents the receiving beamforming vector at the BS to decode stream $s_{2}$. Consequently, the achievable rate to decode the stream $s_{12}$ of UE-1 at BS can denoted as,
\begin{equation}
R_{12}(\textbf{P},\textbf{W},\boldsymbol{\Phi})=\log_2\left(1+\frac{p_{12}|\textbf{w}_{12}^H\left(\textbf{h}_{1,b}+\textbf{h}_{r,b}\boldsymbol{\Phi}\textbf{h}_{r,1}\right)|^2}{\sigma^2}\right),
\end{equation}
where $\textbf{w}_{12} \in \mathbb{C}^{N \times 1}$ represents the receiving beamforming vector at the BS to decode stream $s_{12}$. Finally, the total achievable rate of UE-1 and the sum rate of the system can be, respectively, denoted as,
 \begin{equation}
     R_1=R_{11}+R_{12},
 \end{equation}
 \begin{equation}
     R=R_1+R_2.
 \end{equation}

\subsection{Problem Formulation}
Our objective is to maximize the network sum rate by jointly optimizing the scattering matrix of group-connected BD-RIS, receiving beamforming vectors at BS, and transmission power of UEs while maintaining the rate requirements of each UE. Hence, our sum rate maximization problem can be formulated as follows,
\allowdisplaybreaks
\begin{subequations}
\label{prob:P1}
\begin{flalign}
\centering
 &\mathcal{P}: \max_{\substack{\textbf{W}, \,\boldsymbol{P}, \,\boldsymbol{\Phi}}} \quad \quad R(\textbf{P},\textbf{W},\boldsymbol{\Phi}),\:\label{c1} \\
 &\text{s.t.} \quad p_{11}+p_{12} \le P_{max,1}, p_2 \le P_{max,2}, \label{c3}\\
   &\quad \quad R_{1} \ge R_{th,1}, R_{2} \ge R_{th,2},\label{c8}\\
    &\quad \quad ||\textbf{w}_{11}||=1, ||\textbf{w}_{12}||=1, ||\textbf{w}_{2}||=1,\label{c5}\\
    & \quad \quad \boldsymbol{\Phi}^H_g\boldsymbol{\Phi}_g=\textbf{I}_{M_g}, \boldsymbol{\Phi}_g=\boldsymbol{\Phi}_g^T, g \in {G},\label{c9}\\
    & \quad \quad \boldsymbol{\Phi}=\textrm {blkdiag}(\boldsymbol{\Phi}_1,....,\boldsymbol{\Phi}_G),\label{c7}
\end{flalign}
\end{subequations}
where $\textbf{W}=[\textbf{w}_{11}, \textbf{w}_{12}, \textbf{w}_2]$, $\textbf{P}=[p_{11}, p_{12}, p_2]$. \eqref{c3} denotes the transmission power budget constraints and $P_{max,1}$ and $P_{max,2}$ represent the maximum power budget of for UE-1 and UE-2 respectively. Meanwhile, \eqref{c8} represents the rate requirement constraints for UE-1 and UE-2, respectively. \eqref{c5} represents the  constraint for receiving beamforming vectors. Finally, \eqref{c9}-\eqref{c7} represent the constraints for scattering matrix constraint for group-connected BD-RIS and $M_g$ represents the number of RIS elements in each group.
\section{Solution approach: A Gradient-based Meta learning}
In this section, we describe our gradient-based meta learning approach for solving our formulated optimization problem. 
In conventional DL-based approaches, the $\textbf{W}$, $\textbf{P}$ and $\boldsymbol{\Phi}$ are provided as input for neural networks and then it outputs the target matrices $\textbf{W}^*$, $\textbf{P}^*$, and $\boldsymbol{\Phi}^*$. However, it is quite difficult to understand the procedure inside the black-box neural network. Recently, gradient-based meta-learning method has emerged as a promising approach to handle large-scale optimization problems. In this gradient-based approach, we feed the gradients $\Delta_{\textbf{W}}R$, $\Delta_{\textbf{P}}R$, and $\Delta_{\boldsymbol{\Phi}}R$  as an input the neural networks and which then outputs $\Delta{\textbf{W}}$,$\Delta{\textbf{P}}$, and $\Delta{\boldsymbol{\Phi}}$. These output gradients are added to the initialized or updated $\textbf{W}$,$\textbf{P}$, $\boldsymbol{\Phi}.$ In comparison to the conventional DL approach, this approach offers improved interpretability. 
\subsection{Gradient-based Meta Learning Architecture}
The conventional data-driven meta-learning approach requires substantial offline pre-training with further online adaptation, such as model-agnostic meta-learning. However, these methods do not perform well in variation of data distribution. Moreover, large-scale pre-training and frequent adaptation require substantial energy consumption, making this approach impractical for latency-sensitive and dynamic environments. To tackle these issues, similar to \cite{10623434}, we adopt a pre-training-free, model-driven meta-learning framework, which tries to find the trajectory rather than individual variables. The proposed algorithm consists of three-layer nested cyclic optimization, which includes inner iterations, epoch iterations, and outer iterations. The details of the proposed approach are provided below:
\subsubsection{Inner Iteration}
This layer focuses on optimizing the target variables. It includes three separate networks designed to optimize $\textbf{W}$, $\textbf{P}$, and $\boldsymbol{\Phi}$, referred to as the receiving beamforming network (RBN), transmission power network (TPN), and scattering matrix network (SMN), respectively. In each of these networks, $\textbf{W}$, $\textbf{P}$, and $\boldsymbol{\Phi}$ are updated one after another in each inner iteration. In each network, optimization of the target variable starts from the initial value, while the other variables are inherited from their corresponding network. The target variables are optimized alternatively in their corresponding network until convergence.  Hence, the update process in the $j$-th outer iteration is formulated as,
\begin{equation}
\textbf{W}^*=RBN(\textbf{P}^*,\textbf{W}^{(i,j)},\boldsymbol{\Phi}^*),
\end{equation}
\begin{equation}  \textbf{P}^*=TPN(\textbf{P}^{(i,j)},\textbf{W}^*,\boldsymbol{\Phi}^*),
\end{equation}
\begin{equation} 
\boldsymbol{\Phi}^*=SMN(\textbf{P}^*,\textbf{W}^*,\boldsymbol{\Phi}^{(i,j)}),
\end{equation}
where $\textbf{W}^{(i,j)}$, $\textbf{P}^{(i,j)}$ and $\boldsymbol{\Phi}^{(i,j)}$ represent $\textbf{W}$, $\textbf{P}$, and $\boldsymbol{\Phi}$ in the $i$-th inner iteration of the $j$-th outer iteration. The detail of each neural network is provided below.

\textit{\textbf{Receiving beamforming network (RBN)}}:
In order to optimize the receiving beamforming vectors $\textbf{W}$ in $i$-th inner iteration and $j$-th outer iteration, we write the objective function as $R({\hat{\textbf{P}},\textbf{W}^{(i,j)},\hat{\boldsymbol{\Phi}}}),\label{rate1}$
where $\hat{p}_{11}, \hat{p}_{12},\hat{p}_2$ and $\hat{\boldsymbol{\Phi}}$ represent either the initialized or updated transmission power of UEs and scattering matrix. At first, the sum rate is first computed and the gradient of $\textbf{w}^{(i,j)}$ with respect to sum rate
is fed into the neural network, then the output $\Delta \textbf{w}^{(i,j)}$ is added to $\textbf{w}^{(i,j)}$ which can be given as follows,
\begin{equation}
\begin{split}
&\textbf{w}_{11}^{(i+1,j)}=\textbf{w}_{11}^{(i,j)}+\Delta\textbf{w}_{11}^{(i,j)}, 
\textbf{w}_{12}^{(i+1,j)}=\textbf{w}_{12}^{(i,j)}+\Delta\textbf{w}_{12}^{(i,j)}, \\
&\textbf{w}_{2}^{(i+1,j)}=\textbf{w}_{2}^{(i,j)}+\Delta\textbf{w}_{2}^{(i,j)}. \label{w11}
\end{split}
\end{equation}

\textit{\textbf{Transmission power network (TPN)}}: We denote objective in the $i$-th inner iteration and $j$-th outer iteration as $R_{\textbf{P}}^{(i,j)}$, which can be written as,
$R(\textbf{P}^{(i,j)},\hat{\textbf{W}},\hat{\boldsymbol{\Phi}}).\label{rate2}$
Consequently, we can obtain the computed gradient from TPN, which can later be added to the $p^{(i,j)}$, which can be given as follows,
 \begin{equation}
 \begin{split}
&{p}_{11}^{(i+1,j)}={p}_{11}^{(i,j)}+\Delta{p}_{11}^{(i,j)}, 
{p}_{12}^{(i+1,j)}={p}_{12}^{(i,j)}+\Delta{p}_{12}^{(i,j)}, \\
&{p}_{2}^{(i+1,j)}={p}_{2}^{(i,j)}+\Delta{p}_{2}^{(i,j)}. \label{p2}
\end{split}
 \end{equation}

\textit{\textbf{Scattering matrix network (SMN)}}:
We denote the objective in the $i$-th inner iteration and $j$-th outer iteration as,
$R(\hat{\textbf{P}},\hat{\textbf{W}},\boldsymbol{\Phi}^{(i,j)}).\label{rate3}$
Even though the formulation is similar to the previous networks, the behavior of the target variable involves a trigonometric function, which is quite different from the previous variables and imposes more optimization challenges. Particularly if we update the scattering matrix of BD-RIS directly as $\boldsymbol{\Phi}^{(i+1,j)}=\boldsymbol{\Phi}^{(i,j)}+\Delta \boldsymbol{\Phi}^{(i,j)}$, the change may not be monotonic due to the periodicity of trigonometric functions. As a result, the output of the neural network may exceed its intended range, leading to variations in sum rate and making it challenging to achieve convergence. This may lead to instability during optimization. Given that the period of a trigonometric function is $2\pi$, a customized regulator is utilized to ensure that $\Delta \boldsymbol{\Phi}$ remains confined within the range of $0$ to $2\pi$, which can be given as $\Delta  \tilde{\theta} =\alpha . \delta (\Delta \theta)$, where $\alpha$ acts as a regulatory operator and $\delta(.)$ represents the sigmoid function. We perform this operation for every element of the scattering matrix. Thus, the updated scattering matrix can be obtained as follows,
\begin{align}
\footnotesize
&\boldsymbol{\Phi}_g^{(i+1,j)}= \nonumber \\
\footnotesize
&\begin{bmatrix}
e^{j\theta_{m_g}^{(0,0)}+\Delta \tilde{\theta}_{m_g}} & e^{j\theta_{m_g}^{(0,1)}+\Delta \tilde{\theta}_{m_g}}  & \dots & e^{j\theta_{m_g}^{(0,m_g)}+\Delta \tilde{\theta}_{m_g}} \\
e^{j\theta_{m_g}^{(1,0)}+\Delta \tilde{\theta}_{m_g}} & e^{j\theta_{m_g}^{(1,1)}+\Delta \tilde{\theta}_{m_g}}  & \dots & e^{j\theta_{m_g}^{(1,m_g)}+\Delta \tilde{\theta}_{m_g}} \\
\vdots & \vdots  & \ddots & \vdots \\
e^{j\theta_{m_g}^{(m_g,0)}+\Delta \tilde{\theta}_{m_g}} & e^{j\theta_{m_g}^{(m_g,1)}+\Delta \tilde{\theta}_{m_g}}  & \dots & e^{j\theta_{M_g}^{(M_g,M_g)}+\Delta \tilde{\theta}_{M_g}}
\end{bmatrix},
\end{align}
\begin{equation}
\boldsymbol{\Phi}^{(i+1,j)}=\textrm {blkdiag}(\boldsymbol{\Phi}_1^{(i+1,j)},....,\boldsymbol{\Phi}_G^{(i+1,j)}).
\end{equation}
\subsubsection{Outer iteration}
In each outer iteration, $N_i$ inner iterations contribute to accumulating the meta loss. The meta loss function must encapsulate both the objective (to maximize the sum-rate) and the constraints (to ensure feasibility). We can model the meta loss function as follows,
\allowdisplaybreaks
\begin{subequations}
\label{prob:P1}
\begin{flalign}
\centering
&\mathcal{L}^j=\mathcal{L}_{rate}+\mathcal{L}_{threshold}+\mathcal{L}_{norm}+\mathcal{L}_{RIS}, \label{m}\\
&\mathcal{L}_{rate}=-(R_1+R_2), \label{1}\\
&\mathcal{L}_{threshold}=\lambda (\Xi_1+\Xi_2),\mathcal{L}_{norm}=\zeta (\Gamma_1+\Gamma_2+\Gamma_3)
\label{3},\\
&\mathcal{L}_{RIS}=\eta{(\Sigma_1+\Sigma_2)},\mathcal{L}_{power}= \mu(\Upsilon_1+\Upsilon_2),\label{5}\\
&\lambda  = \begin{cases}0, & \mbox{if } \mbox{$\Xi_1, \Xi_2$}, \\  1, & \mbox{if } \mbox{otherwise}, \end{cases}
\zeta = \begin{cases}0, & \mbox{if } \mbox{$\Gamma_1, \Gamma_2, \Gamma_3$}, \\  1, & \mbox{if } \mbox{otherwise}, \end{cases}\\
&\eta= \begin{cases}0, & \mbox{if } \mbox{$\Sigma_1,\Sigma_2$}, \\  1, & \mbox{if } \mbox{otherwise}, \end{cases}\mu= \begin{cases}0, & \mbox{if } \mbox{$\Upsilon_1, \Upsilon_2$}, \\  1, & \mbox{if } \mbox{otherwise}, \end{cases}
\end{flalign}
\end{subequations}
where $\Xi_1=(R_{th,1}-R_1) \le 0, \Xi_2=(R_{th,2}-R_2) \le 0$,
$\Gamma_1=||W_{11}-1||_2 \le 0,\Gamma_2=||W_{12}-1||_2 \le 0,\Gamma_3=||W_{2}-1||_2 \le 0$, $\Sigma_1 =\sum_{g \in G}||\boldsymbol{\Phi}_g^H\boldsymbol{\Phi}_g-\boldsymbol{I}_{M_g}||_2 \le 0$, $\Sigma_2=||\boldsymbol{\Phi}_g-\boldsymbol{\Phi}_g^T||_2 \le 0$, $\Upsilon_1=(p_{11}+p_{12})-{P_{max,1}} \le 0, \Upsilon_2 =p_2-P_{max,2} \le 0$. Furthermore, $\lambda$, $\zeta$, $\eta$, and $\mu$ represent the regularization parameter to balance enforcing the rate constraints and maximizing the sum rate.
Meanwhile, the first term in \eqref{3} is introduced to replace the rate constraints of \eqref{c8}.  Meanwhile, the second term in \eqref{3} replaces the constraints of \eqref{c5}. The first term of \eqref{5} replaces the constraints in \eqref{c5} and the second term of \eqref{5} replaces the constraints in \eqref{c3}. Consequently, meta loss function is penalized if the violation of constraints occurs.
\subsubsection{Epoch iteration} 
This iteration is responsible for updating the parameters of neural networks. For each epoch iteration, there are $N_o$ outer iterations. Once $N_o$ outer iterations are performed,the losses are summed and averaged as
\begin{equation}
    \bar{\mathcal{L}}=\frac{1}{N_o}\sum_{j=1}^{N_o}\mathcal{L}^j.
\end{equation}
Then, backward propagation is conducted and the Adam optimizer is used to update the neural networks embedded in these networks, as depicted below:
\begin{equation}
\theta_{\textbf{W}}^*=\theta_{\textbf{W}}+\beta_{\textbf{W}}.\textrm{Adam}(\Delta_{\theta_{\textbf{W}}}\mathcal{\bar{L}},\theta_{\textbf{W}}),\label{w}
\end{equation}
\begin{equation}
\theta_{\textbf{P}}^*=\theta_{\textbf{P}}+\beta_{\textbf{P}}.\textrm{Adam}(\Delta_{\theta_{\textbf{P}}}\mathcal{\bar{L}},\theta_{\textbf{P}}),\label{p}
\end{equation}
\begin{equation}
\theta_{\boldsymbol{\Phi}}^*=\theta_{\boldsymbol{\Phi}}+\beta_{\boldsymbol{\Phi}}.\textrm{Adam}(\Delta_{\theta_{\boldsymbol{\Phi}}}\mathcal{\bar{L}},\theta_{\boldsymbol{\Phi}}),\label{phi}
\end{equation}
where $\beta_{\textbf{W}}$, $\beta_{\textbf{P}}$, and $\beta_{\boldsymbol{\Phi}}$ represent the learning rates of the three networks, respectively. The proposed gradient-based meta-learning is provided in Algorithm 1. 

\textit{Complexity Analysis}: The proposed gradient-based meta-learning algorithm does not involve any matrix inversion which decreases its complexity. Following \cite{10623434}, we can calculate the complexity our proposed algorithm as below. $N_eN_oN_i O(K^2(M^2+MN)+K^2(M^2+MN)+K^2(M^2+MN))$,where $K$ is the number of UEs.
\begin{algorithm}[!t]
\footnotesize
\caption{Proposed algorithm}\label{alg:3}
 Randomly initialize $\theta_{\textbf{P}},\theta_{\textbf{W}},\theta_{\boldsymbol{\Phi}},\textbf{P}^{(0,1)}$, $\textbf{W}^{(0,1)},\boldsymbol{\Phi}^{(0,1)}$;\\
 Initialize inner iterations, $N_i$, outer iterations, $N_o$, and epoch iterations, $N_e$;\\
\For {$e={1,2,...,N_e}$}{
$\bar{\mathcal{L}}=0$;\\
MAX$=0$;\\
\For {$j=1,2,...N_o$}{
  $\textbf{W}^{(0,j)}=\textbf{W}^{(0,1)}$ ;\\
   $\textbf{P}^{(0,j)}=\textbf{P}^{(0,1)}$;\\
    $\boldsymbol{\Phi}^{(0,j)}=\boldsymbol{\Phi}^{(0,1)}$;\\
      \For {$i= 1,2,...,N_i$}{
      $R_{\textbf{W}}^{(i-1,j)}=R(\textbf{W}^{(i-1,j)},\textbf{P}^*,\boldsymbol{\Phi}^*)$;\\
      $\Delta \textbf{W}^{(i-1,j)}$= $\textrm{RBN}(\Delta_{\textbf{W}}R_{\textbf{W}}^{(i-1,j)})$;\\
      $\textbf{W}^{(i,j)}=\textbf{W}^{(i-1,j)}+\Delta \textbf{W}^{(i-1,j)}$;
      }
      $\textbf{W}^*=\textbf{W}^{(N_i,j)}$;\\
        \For {$i= 1,2,...,N_i$}{
      $R_{\textbf{P}}^{(i-1,j)}=R(\textbf{P}^{(i-1,j)},\textbf{W}^*,\boldsymbol{\Phi}^*)$;\\
      $\Delta \textbf{P}^{(i-1,j)}$= $\textrm{TPN}$
      $(\Delta_{\textbf{P}}R_{\textbf{P}}^{(i-1,j)})$;\\
      $\textbf{P}^{(i,j)}=\textbf{P}^{(i-1,j)}+\Delta \textbf{P}^{(i-1,j)}$;
      }
      $\textbf{P}^*=\textbf{P}^{(N_i,j)}$;\\
        \For {$i= 1,2,...,N_i$}{
      $R_{\boldsymbol{\Phi}}^{(i-1,j)}=R(\boldsymbol{\Phi}^{(i-1,j)},\textbf{W}^*,\textbf{P}^*)$;\\
      $\Delta \boldsymbol{\Phi}^{(i-1,j)}$= $\textrm{SMN}$
      $(\Delta_{\boldsymbol{\Phi}}R_{\boldsymbol{\Phi}}^{(i-1,j)})$;\\
      $\boldsymbol{\Phi}^{(i,j)}=\boldsymbol{\Phi}^{(i-1,j)}+\Delta \boldsymbol{\Phi}^{(i-1,j)}$;
      }
    $\boldsymbol{\Phi}^*=\boldsymbol{\Phi}^{(N_i,j)}$;\\
   Calculate $\mathcal{L}^j$ as \eqref{m};\\
   $\bar{\mathcal{L}}=\bar{\mathcal{L}}+\mathcal{L}^j$;\\
   \textbf{If} {$-\mathcal{L}^j>MAX$}\\
        \quad MAX=$-\mathcal{L}^j$;\\
        \quad $\textbf{W}_{opt}=\textbf{W}^*$;\\
        \quad $\textbf{P}_{opt}=\textbf{P}^*$;\\
        \quad $\boldsymbol{\Phi}_{opt}=\boldsymbol{\Phi}^*$;\\
   \textbf{end if}
   }
   
$\bar{\mathcal{L}}=\frac{1}{N_o}*\bar{\mathcal{L}}$;\\
update $\theta_{\textbf{P}}$ as \eqref{p}, $\theta_{\textbf{W}}$ as \eqref{w}, \\
\textbf{If} {$(e+1) \mod 5==0$}\\
\quad $\theta_{\boldsymbol{\Phi}}$ as \eqref{phi}\\
\textbf{end if}
}
return $\textbf{W}_{opt},\textbf{P}_{opt},\boldsymbol{\Phi}_{opt}$
\end{algorithm}
\section{Numerical Results}
In this section, we evaluate the performance of our proposed approach through simulation results. The locations of BS, BD-RIS, UE-1, and UE-2 can be given as $(0,0,10), (100,2,10),(80,0,0),(110,0,0)$. For large-scale fading, the distance-based path loss model is considered which is $PL_x=L_o(d_x/d_o)^{-\eta}$ where $L_o$ represents the PL exponent at reference distance $d_o$, $\eta$ denotes the PL exponent and $d_x$ represents the distance between nodes in the $x$-th dimension. The cascaded link between BS-BD-RIS-UE is modeled as a Rician fading which is modeled as follows:
\begin{align}
    \boldsymbol{h}_x= \sqrt{PL_x}\left(\sqrt{\left(\frac{1}{1+k_x}\right)}g_x+\sqrt{\left(\frac{k_x}{1+k_x}\right)}\hat{g}_x\right),
\end{align} 
where $k_x$ represents the rician factor, $g_x$ represents the LoS component and $\hat{g}_x$ represents the non LoS component. The simulation parameters and neural network parameters are provided in Table I and Table II respectively. We run the simulations in Google Colab with PyTorch 2.4.1 and Python 3.10.12. Learning rates of $\beta_{\textbf{W}}$ and $\beta_{\textbf{P}}$ are $1e^{-3}$ and for $\beta_{\boldsymbol{\Phi}}$ is $1.5e^{-3}$. The value for all regularization parameters has been varied from 0.1 to 0.0001.
\par
\begin{figure}
    \centering
\includegraphics[width=0.85\columnwidth]{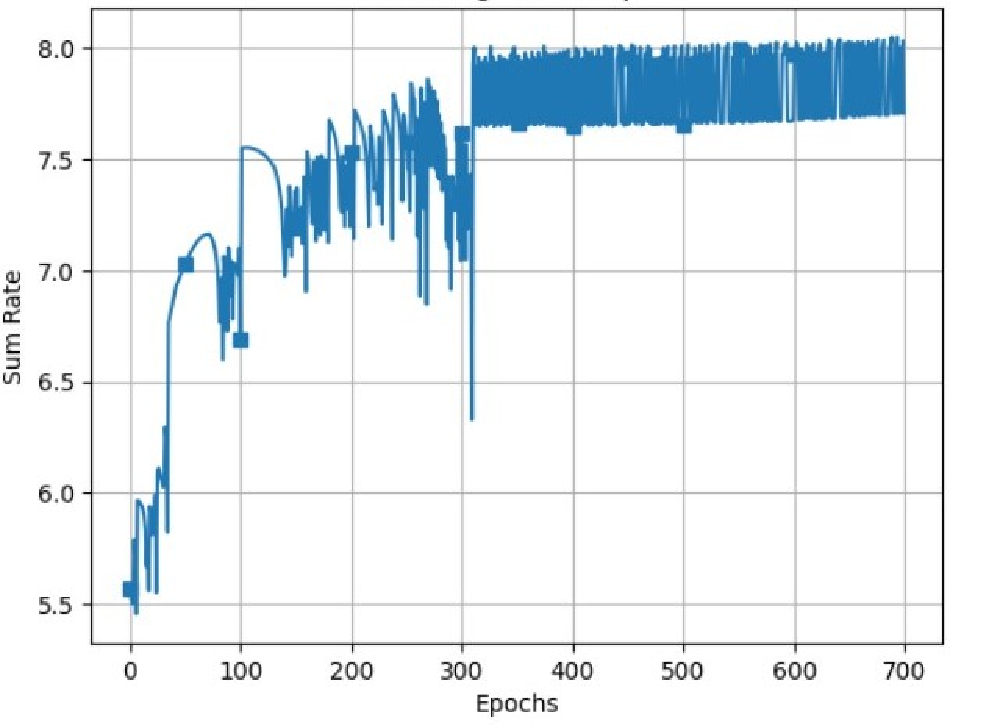}
\caption{Convergence of proposed algorithm for $N=4$}
\label{convregence1}
\end{figure}

\begin{figure}
    \centering
\includegraphics[width=0.9\columnwidth]{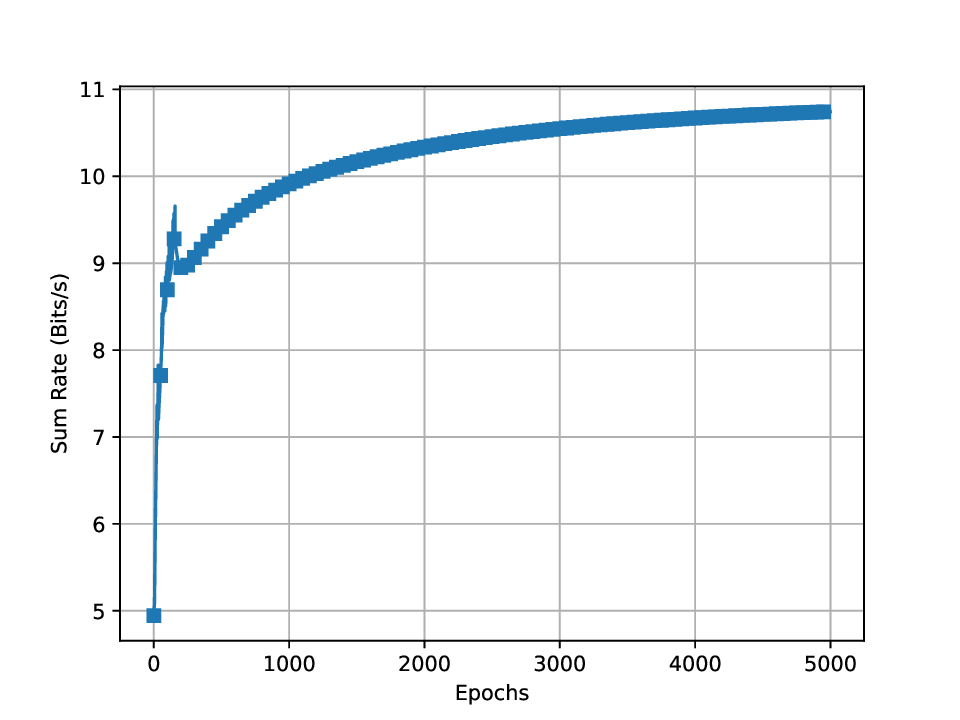}
\caption{Convergence of proposed algorithm for $N=36$ }
\label{convergence2}
\end{figure}

 Fig. \ref{convregence1} shows the convergence behaviour of our proposed scheme for a single channel realization and under the setting of parameters in Table I. Fig. \ref{convergence2} also shows the convergence behaviour but for $N=36$. It can be seen from Fig. \ref{convregence1} that convergence speed is very high and the proposed algorithm converges after 300 epochs. Meanwhile, when we increase the number of antennas to 36, the convergence is reached after 3000 epochs. This implies that when the number of variables increases, it also impacts the convergence speed as well. However, interestingly, in Fig. \ref{convregence1} when $N=4$, the algorithm seems to struggle to find the trajectory to reach the convergence point. On the other hand, when $N=36$ in Fig. \ref{convergence2}, the algorithm seems to be much smoother to find the convergence. This behaviour can be explained as follows. When the number of variables in the optimization problem is of moderate size, the feasibility region to find the initial starting point for the trajectory to reach the optimal point is minimal, which leads to lots of violations of the constraints. On the other hand, when the number of variables in the optimization problem is large enough, the feasibility region to find the trajectory also becomes large enough which makes it easier to find a good starting point for faster adaptation \cite{andrychowicz2016learning}. This behaviour of the proposed algorithm makes it more suitable to solve large-scale optimization problems. 
\par
\begin{figure}
    \centering
\includegraphics[width=0.9
\columnwidth]{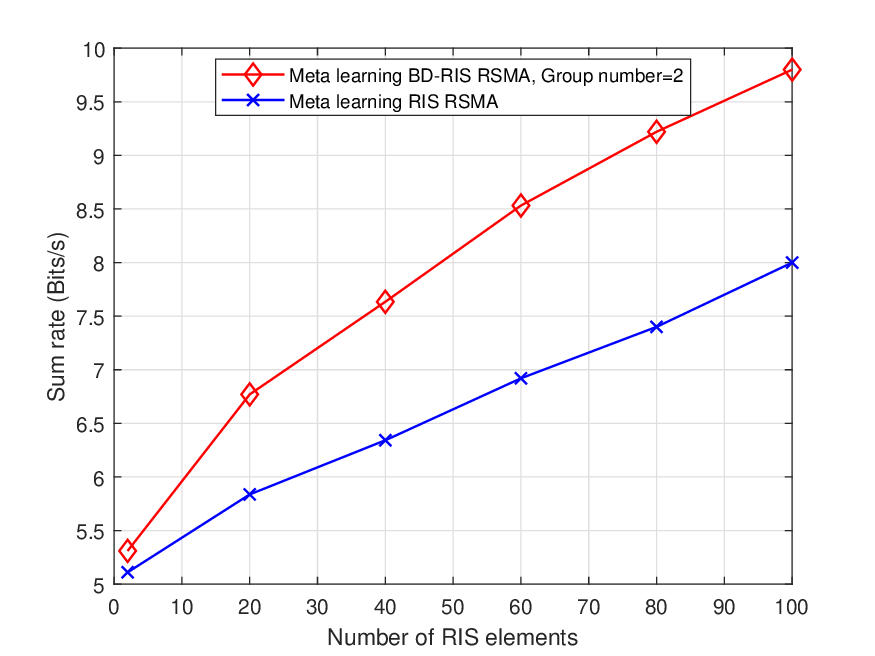}
\caption{Sum rate vs. number of RIS elements, $N=4$}
\label{elements}
\end{figure}

 Fig. \ref{elements} illustrates the performance of BD-RIS with the proposed gradient-based meta learning while we vary the number of RIS elements from 2 to 100. In our group-connected BD-RIS structure, we have taken $G=2$, which means we have divided our BD-RIS into two groups, and each group consists of half the RIS elements. We compare our proposed scheme with the conventional RIS RSMA with meta-learning framework. It can be seen from the figure that when we increase the number of RIS elements, the sum rate for both of the schemes tends to increase as well. However, BD-RIS RSMA can perform better than RIS-RSMA. It is because BD-RIS can leverage much higher flexibility within a group; all elements of the scattering matrix can work together to project the signal direction toward the destination. On the other hand, in the case of RIS-RSMA, only the diagonal elements of the phase shift matrix work together to direct the signal. We can see from the figure that at $M=100$, we can achieve around 22.5$\%$ improvement over conventional RIS.
\par
\begin{figure}
    \centering
\includegraphics[width=0.9\columnwidth]{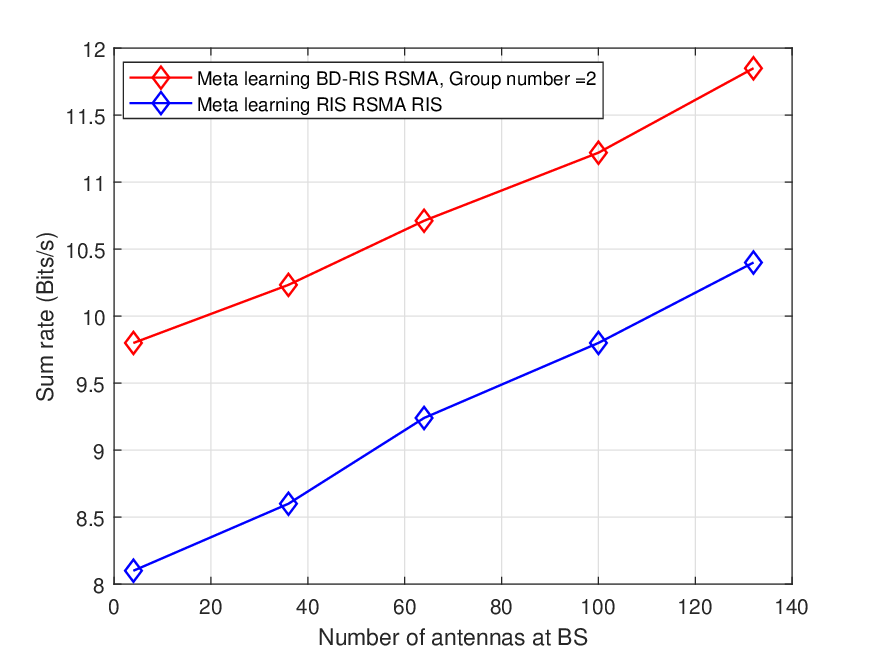}
\caption{Sum rate vs. number of antennas at BS, $M=100$}
\label{antennas}
\end{figure}

 Fig. \ref{antennas} presents the sum rate while varying the number of antennas at BS. We vary the antennas at BS from 4 to 132 while keeping the number of RIS elements $M=100$. It can be seen from the figure that sum rate of both of the schemes increases when we increase the number of antennas. At $N=132$, BD-RIS RSMA can achieve around 13.4\% improvement over RIS RSMA. It can be noticeable from figure that increasing the number of antennas at BS can improve the spatial diversity of both schemes.
\begin{table}
\centering
\caption{Simulation parameters}
\begin{tabular}{ |l|c| }
 \hline
  \textbf{Parameter}& \textbf{Value} \\
 \hline
 $R_{th,1}, R_{th,2}$ & 1 bps\\
 \hline
 $k_x, L_o, \sigma^2$ & 5 dB, -30 dB, -80 dBm\\
 \hline
 $P_{max,1}$, $P_{max,2}$ & 23 dBm\\
 \hline
\end{tabular}
\end{table}
\begin{table}
\centering
\caption{Number of neurons in the neural networks}
\begin{tabular}{ |l|c|c|c| }
 \hline
  \textbf{Layer name} & \textbf{RBN} & \textbf{TPN} & \textbf{SMN}\\
  \hline
Input Layer & 2 & 1 & $M_g$ \\
 \hline
Linear Layer  & $200$ & 200 & 200 \\
 \hline
 ReLU Layer & 200 & 200 & 200\\
 \hline
Output Layer & 2& 1 & $M_g$\\
 \hline
\end{tabular}
\end{table}
\section{Conclusion}
In this paper, we propose a pre-training free algorithm for BD-RIS-aided uplink RSMA communication. We jointly optimize the receiving beamforming vectors at BS, scattering matrix of BD-RIS and transmission power of UEs. We utilize gradient-based meta learning to solve our optimization problem which can solve large-scale optimization problems  with low complexity and higher speed of convergence. Simulation results demonstrate that our proposed scheme outperforms conventional RIS-RSMA scheme. As a future work, we intend to extend this work with large number of UEs under both perfect and imperfect CSI.
\bibliographystyle{IEEEtran}
\bibliography{IEEEabrv,biblio}
\vfill
\end{document}